\documentclass[longauth,structabstract]{aa} 
\usepackage{graphicx}
\usepackage{color}
\usepackage{natbib}
\bibpunct{(}{)}{;}{a}{}{,} 
\newcommand{\nc}{\newcommand}
\nc{\lsun}{\ensuremath{\mathrm{L}_\odot}}
\nc{\msun}{\ensuremath{\mathrm{M}_\odot}}
\nc{\tex}{\ensuremath{\mathrm{T}_{\rm ex}}}
\nc{\cthree}{C$_3$}
\nc{\cthreehtwo}{$c$-C$_3$H$_2$}
\nc{\kms}{\mbox{km\,s$^{-1}$}}
\nc{\Kkms}{\mbox{K\,km\,s$^{-1}$}}
\nc\micron{\mbox{$\mu$m}}
\nc{\Trot}{$T_{\rm rot}$}%
\nc{\Ntot}{$N(C_3)$}%
\nc{\Tc}{$T_{\rm c}$}%
\nc{\Tdust}{$T_{\rm dust}$}%
\nc{\Tex}{$T_{\rm ex}$}%
\nc{\Tkin}{$T_{\rm kin}$}%
\nc{\cmcub}{\mbox{cm$^{-3}$}}
\nc{\cmsq}{\mbox{cm$^{-2}$}}
\newcommand{\HII}{H {\sc ii}}
\newcommand{\vlsr}{$v_{\rm LSR}$}
\newcommand\arcdeg{\mbox{$^\circ$}}%

\begin{document}
   \title{
  The chemistry of C$_3$ \& Carbon Chain Molecules in DR21(OH)
    \thanks{Herschel is an ESA space observatory
    with science instruments provided by European-led Principal
    Investigator consortia and with important participation from NASA.}
}

   \author{
B.~Mookerjea\inst{\ref{tifr}}, 
G.~E.~Hassel\inst{\ref{siena}},
M. Gerin\inst{\ref{ens}},
T.~Giesen\inst{\ref{kosma}}, 
J.~Stutzki\inst{\ref{kosma}}, 
E.~Herbst\inst{\ref{ohio}}, 
J.~H.~Black\inst{\ref{chalmers}}, 
P.~F.~Goldsmith\inst{\ref{jpl}},
K.~M.~Menten \inst{\ref{mpifr}},
J.~Kre{\l}owski\inst{\ref{torun}}, 
M.~De Luca\inst{\ref{ens}},
T.~Csengeri \inst{\ref{mpifr}},
C.~Joblin \inst{\ref{univtoulouse},\ref{cnrstoulouse}}, 
M.~Ka{\'z}mierczak\inst{\ref{sron}}, 
M.~Schmidt\inst{\ref{ncacpoland}}
J.~R.~Goicoechea \inst{\ref{cantab}}, 
J.~Cernicharo\inst{\ref{cantab}}
}

\institute{Tata Institute of Fundamental Research, Homi Bhabha Road,
Mumbai 400005, India \email{bhaswati@tifr.res.in}\label{tifr} 
\and 
Department of Physics and Astronomy, Siena College, Loudonville, NY
12211, USA\label{siena}
\and LERMA, CNRS, Observatoire de Paris and ENS, France \label{ens}
\and
I. Physikalisches Institut, University of Cologne, Germany\label{kosma}
\and Depts.\ of Physics, Astronomy \& Chemistry, Ohio State Univ. USA.\label{ohio}  
\and  Onsala Space Observatory, Chalmers University of Technology, SE-43992 Onsala, Sweden \label{chalmers} 
\and JPL, California Institute of Technology, Pasadena, USA\label{jpl}
\and MPI f\"ur Radioastronomie, Bonn, Germany \label{mpifr}.
\and Nicolaus Copernicus University, Toru{\'n}, Poland\label{torun}
\and
Universit\'e de Toulouse; UPS-OMP; IRAP;  Toulouse,
France\label{univtoulouse}
\and
CNRS; IRAP; 9 Av. colonel Roche, BP 44346, F-31028 Toulouse cedex 4, France
\label{cnrstoulouse}
\and 
SRON Netherlands Institute for Space Research, Landleven 12, 9747 AD
Groningen, The Netherlands\label{sron}
\and  Nicolaus Copernicus Astronomical Center (CMAK), Toru{\'n},
Poland\label{ncacpoland}.
\and
Centro de Astrobiolog\'{\i}a, CSIC-INTA, 28850, Madrid, Spain\label{cantab}
}

  \date{Received \ldots accepted \ldots}

\abstract
{\cthree\ is the smallest pure carbon chain detected in the dense
environment of star forming regions, although diatomic C$_2$ is detected
in diffuse clouds.  Measurement of the abundance of \cthree\ and the
chemistry of its formation  in dense star forming regions has remained
relatively unexplored.}
{We aim to identify the primary \cthree\ formation routes in dense star
forming regions  following a chemical network producing species like
CCH and \cthreehtwo\ in the star forming cores associated
with DR21(OH), a high mass star forming region.  }
{We have observed velocity resolved spectra of four ro-vibrational
far-infrared transitions of \cthree\ between the vibrational ground
state and the low-energy $\nu_2$ bending mode  at frequencies between
1654--1897\,GHz using HIFI on board {\em Herschel}, in DR21(OH).
Several transitions of CCH and \cthreehtwo\ have also
been observed with HIFI and the IRAM 30m telescope.  
Rotational temperatures and column densities for all chemical species
were estimated.  A gas and grain warm-up model was used to obtain
estimates of  densities and temperatures of the envelope.  The chemical
network in the model has been used to identify the primary \cthree\
forming reactions in DR21(OH).}
{We have detected \cthree\ in absorption in four far-infrared
transitions, $P(4)$, $P(10)$, $Q(2)$ and $Q(4)$. The continuum sources
MM1 and MM2 in DR21(OH) though spatially unresolved, are sufficiently 
separated in
velocity to be identified in the \cthree\ spectra. All \cthree\
transitions are detected from the embedded source MM2 and the
surrounding envelope, whereas only $Q(4)$ \& $P(4)$ are detected toward
the hot core MM1. The abundance of \cthree\ in the envelope and MM2 is
$\sim 6\times10^{-10}$ and $\sim 3\times10^{-9}$ respectively.  For CCH and
\cthreehtwo\ we only detect emission from the envelope and MM1.  The
observed CCH, \cthree\, and \cthreehtwo\  abundances are most consistent
with a chemical model with $n_{\rm H_2}\sim$ 5$\times 10^{6}$~\cmcub,
post-warm-up dust temperature, $T_{\rm max}$=30~K and a time of $\sim$
0.7--3~Myr.}  {Post warm-up gas phase chemistry of CH$_4$ released from
the grain at $t \sim$ 0.2\,Myr and lasting for 1\,Myr can explain the
observed \cthree\ abundance in the envelope of DR21(OH) and no mechanism
involving photodestruction of PAH molecules is required. The chemistry
in the envelope is similar to the warm carbon chain chemistry (WCCC)
found in lukewarm corinos.  We interpret the observed lower \cthree\
abundance in MM1 as compared to MM2 and the envelope to be due to the
destruction of \cthree\ in the more evolved MM1.  The timescale for the
chemistry derived for the envelope is consistent with the dynamical
timescale of 2\,Myr derived for DR21(OH) in other studies.  }


  \titlerunning{\cthree\ in DR21(OH)}
        \authorrunning{Mookerjea et al.}
   \maketitle

\section{Introduction}

\begin{table*}[t]
\begin{center}
\caption{Properties of the sources in DR21(OH).
\label{tab_srcprop}}
\begin{tabular}{cccrrrrrc}
\hline
\hline
Component & $\alpha$(2000) & $\beta$(2000) & $v_{\rm LSR}$ & Mass$^{a}$ &
$T_{\rm d}$ & N(H$_2$)$^{b}$ & $<n(H_2)>$ & References\\
 & & & (km~s$^{-1}$) & (\msun) & & (10$^{23}$\cmsq) & (10$^7$ \cmcub) &\\
\hline
MM1 & 20$^{\rm h}$39$^{\rm m}$01\fs0 & 42\arcdeg22\arcmin48\arcsec &
$-4.1\pm$0.3 & 88 & 58 & 4.5 & 4.1 & 1,2\\
MM2 & 20$^{\rm h}$39$^{\rm m}$00\fs4 & 42\arcdeg22\arcmin43\farcs8 &
$-0.7\pm$0.3 & 143 & 30  & 7.4 & 6.8 & 1,2\\
Envelope & \ldots & \ldots & $-3.1\pm$0.3 & 118$^{c}$  & & 3.5 & 0.3 & 1,3,4\\
\hline
\end{tabular}
\end{center}

$^{a}$ All estimated for a distance of 1.5~kpc\\
$^{b}$ Within a 15\arcsec\ beam\\
$^{c}$ Mass of the envelope derived by removing the contributions of MM1
and MM2 from the total mass of the region estimated by \citet{motte2007}. \\
References: (1) \citet{mangum1991}, (2) \citet{mangum1992}, (3)
\citet{wilson1990}, (4)\citet{motte2007}.
\end{table*}

\begin{table*}[t]
\begin{center}
\caption{Spectroscopic parameters for the \cthree\ \& CCH transitions observed with 
HIFI/Herschel
\label{tab_specdata}}
\begin{tabular}{lcllrc}
\hline \hline
Species & Transition   & Frequency & {A-coeff} & E$_l$  & Beam Size\\
        &              & (MHz)     & (s$^{-1}$)  &  (cm$^{-1}$) & \\
\hline
\cthree, ($J$,$v$) & (9,1) -- (10,0) P(10) & 1654081.66 & 2.38~10$^{-3}$  & 47.3 & 13\arcsec\\\
               & (3,1) --  (4,0) P(4) & 1787890.57 & 2.72~10$^{-3}$  & 8.6 & 12\arcsec\\\
               & (2,1) --  (2,0) Q(2) & 1890558.06 & 7.51~10$^{-3}$  & 2.6 & 11\arcsec\\\
               & (4,1) --  (4,0) Q(4) & 1896706.56 & 7.58~10$^{-3}$  & 8.6& 11\arcsec\  \\
\hline
CCH, N$_{J,F}$  & 6$_{13/2,7}$ -- 5$_{11/2,6}$  & 523971.5704 & 4.58~10$^{-4}$ & 43 & 40\arcsec \\
                & 6$_{13/2,6}$ -- 5$_{11/2,5}$  & 523972.1630 & 4.53~10$^{-4}$ & 43 & 40\arcsec \\
CCH, N$_{J,F}$  & 6$_{11/2,6}$ -- 5$_{11/2,5}$  & 524033.9075 & 4.51~10$^{-4}$ & 43 & 40\arcsec\\
                & 6$_{11/2,5}$ -- 5$_{9/2,4}$   & 524034.5305 & 4.43~10$^{-4}$ & 43 & 40\arcsec\\

\hline
\end{tabular}
\end{center}
$^\ast$ Herschel HIFI beamsizes are taken from \citet{roelfsema2012}.
\end{table*}
\begin{table*}[t]
\begin{center}
\caption{Spectroscopic \& Observational parameters for the 
transitions of CCH and \cthreehtwo\  observed with IRAM 30~m.
\label{tab_iramspec}}
\begin{tabular}{lcllrccc}
\hline \hline
Species & Transition   & Frequency & {A-coeff} & E$_l$ & $\theta_{\rm
FWHM}$ & F$_{\rm eff}^a$ & B$_{\rm eff}^b$\\
        &              & (MHz)     & (s$^{-1}$)  &  (cm$^{-1}$) & (\arcsec) & & \\
\hline
CCH, N$_{J,F}$ & 1$_{3/2,1}$ -- 0$_{1/2,1}$ &  87284.1050 & 2.60~10$^{-7}$ & 0.0015& 29\arcsec & 0.95 & 0.75\\
            & 1$_{3/2,2}$ -- 0$_{1/2,1}$ &  87316.8980 & 1.53~10$^{-6}$ & 0.0015& 29\arcsec & 0.95 & 0.75\\
            & 1$_{3/2,1}$ -- 0$_{1/2,0}$ &  87328.5850 & 1.27~10$^{-6}$ & 0.0015& 29\arcsec & 0.95 & 0.75\\
\hline
\cthreehtwo, ($J_{K_a},{K_c}$) & 2$_{1,2}$ -- 1$_{0,1}$ & 85338.8930 & 2.32~10$^{-5}$ & 1.7 & 29\arcsec & 0.95 & 0.75\\
\hline
\hline
\end{tabular}
\end{center}
$^a$ $F_{\rm eff}$ Forward efficiency and $^b$ $B_{\rm eff}$  Beam efficiency of the telescope.
\end{table*}

Small carbon chains are important for the chemistry of stellar and
interstellar environments for several reasons: they are ubiquitous
throughout the  interstellar medium \citep{adamkovics2003}, they  are
likely to participate in the formation of long carbon chain molecules,
and they are products of photo-fragmentation cascades of polycyclic
aromatic hydrocarbons (PAHs) \citep{radi1988,pety2005}.  Triatomic
carbon, \cthree, was first tentatively identified in interstellar gas by
\citet{vanorden1995} and \citet{haffner1995}.  The mid-infrared spectrum
of \cthree\ ($\nu_3$ antisymmetric stretching mode) was measured in the
circumstellar envelope of CW Leo (IRC +10216) by \citet{hinkle1988}, and
in low-resolution interstellar absorption  in the far-IR ($\nu_2$
bending mode) toward the Sgr B2 star forming region and IRC+10216 by
\citet{cernicharo2000}.  \citet{giesen2001} discussed new laboratory
data on the vibrational spectrum of \cthree\ in its low-frequency
bending mode and re-visited the first identification of the $\nu_2$
$R(2)$ line in absorption toward Sgr B2 \citep{vanorden1995}. The
abundance and excitation of \cthree\ with a large range of rotational
temperatures in translucent clouds have also been determined
convincingly \citep{maier2001,roueff2002,oka2003,adamkovics2003} through
observations at optical wavelengths. \citet{galazutdinov2002} also
demonstrated that the \cthree\ abundance is related neither to
interstellar reddening nor to the intensities of diffuse interstellar
bands.

DR21(OH) lies 2\arcmin\ North of the \HII\ region DR21, in the Cygnus X
\HII\ complex. The distance to DR21(OH) has recently been accurately
determined by trigonometric parallax measurements of its associated
methanol masers as $1.50^{+0.08}_{-0.07}$~kpc \citep{rygl2011}.
Interferometric and high resolution single-dish observations both in
continuum and molecular lines have shown multiple peaks in DR21(OH)
\citep{wilson1990,mangum1991,mangum1992,chandler1993}.
\citet{mangum1992} resolved the main DR21(OH) peak into two sources, MM1
and MM2, and MM2 into 2 sub-sources, MM2-A and MM2-B.  Synthesis imaging
of NH$_3$ emission from the region has clearly resolved MM1 and MM2,
having radial velocities $-4.1\pm$0.3~\kms\ and $-0.7\pm$0.3~\kms,
respectively \citep{mangum1992}.  Both the star-forming cores MM1 \& MM2
in DR21(OH) are young, having no visible \HII\ region and only very weak
continuum emission at the centimeter wavelengths \citep{argon2000}.
Dust continuum observations suggest that MM1 is the brighter source ($L
= 1.7\times 10^4$~\lsun; ZAMS B0V star) showing evidence of  star
formation, whereas MM2 ($L = 1.2\times 10^3$~\lsun\ ; early B star)
though more massive is fainter and most likely at an earlier stage of
evolution.  The region is characterized as a high-mass star-forming
region due to the detection of centimeter and millimeter maser emission
from numerous transitions e.g., H$_2$O, OH, CH$_3$OH \citep{araya2009,
fish2005, mangum1992, plambeck1990}.  MM1 also contains ground-state OH
masers at 1.6~GHz and 6.7~GHz class II CH$_3$OH masers near the peak of
the dust and centimeter continuum emission
\citep{argon2000,fish2005,rygl2011}.  In a recent interferometric study
of the continuum emission at 1.4~mm from DR21(OH), \citet{zapata2012}
have resolved MM1 and MM2 into 9 compact sources. Five of the compact
sources are associated with MM1 and four with MM2. Two of the compact
sources associated with MM1 (SMA6 and SMA7) seem to show hot core
activity.  \citet{white2010} have observed multiple transitions of CO
from 4--3 to 13--12 using SPIRE/Herschel to conclude the presence of two
components at 80~K and 180~K from a rotation temperature analysis. From
a LVG analysis these authors obtain T$_{\rm kin}$ = 125~K and $n_{\rm
H_2}$ = 7$\times 10^4$~\cmcub.  We summarize selected properties of all
the emission components in DR21(OH) as available from literature in
Table ~\ref{tab_srcprop}.  All values in Table~\ref{tab_srcprop}
correspond to a distance of 1.5~kpc.

The Heterodyne Instrument for the Far-Infrared
\citep[HIFI;][]{deGraauw2010} on board the Herschel Space Observatory
\citep{pilbratt2010}, with its broad frequency coverage, high
sensitivity and spectral resolution has provided for the first time
the opportunity for a systematic study of carbon chain molecules such
as \cthree\ through probing several ro-vibrational lines at full
spectral resolution.  The results presented here are a part of the
PRISMAS (``PRobing InterStellar Molecules with Absorption line
Studies”) Key Program \citep{gerin2010}.  

The primary aim of the paper is to understand the \cthree\ formation
mechanism. Observations of diffuse interstellar gas at optical and IR
wavelengths have shown a strong correlation between the column densities
of \cthree\ and C$_2$. This suggests that \cthree\ and C$_2$ are in the
same chain of chemical reactions \citep{oka2003}.  The formation routes
for \cthree\ starting from C$_2$ also involved the production of CCH and
\cthreehtwo. However no transitions of C$_2$ are available at longer
wavelengths (far-infrared and longer) which can be used to detect C$_2$
in dense star forming regions where \cthree\ is detected in absorption
\citep{mookerjea2010}.  Thus, to understand the formation pathway of
\cthree\ we compare the  abundances of \cthree, CCH, and \cthreehtwo,
with results of chemical models of dense star forming cores. 

\section{Observations}
\subsection{Herschel Data: \cthree\ \& CCH}

We have observed in DR21(OH) four lines of the $\nu_2$ fundamental band,
$P(4)$, $Q(2)$, $Q(4)$ and $P(10)$, of triatomic carbon using the upper
sideband of the HIFI bands 7a, 7b and the lower sideband of band 6b of
the HIFI receiver.  All \cthree\ spectra were observed at a single
position in DR21(OH) with coordinates $\alpha$(J2000) =  20:39:01.00
$\delta$(J2000)= 42:22:48.0, in the dual beam switch mode.  The
observations were carried out on 2010 June 25, and November 25 to 27.
The details of the observations and data reduction are given in
\citet{mookerjea2010}.  All spectra were smoothed to resolution between
$\sim$ 0.16 to 0.18~\kms\ and the rms noise level for the spectra is
0.03 to 0.04~K.  All the observed \cthree\ lines are at frequencies
for which HEB mixers are used in HIFI, and we have noticed a variation
in the level of the continuum observed for different LO settings and
polarizations for a particular spectral line.  In the results presented
here there is only a shift in the continuum level, with little or no
change in the depth of absorption. We have verified that this shift in
the continuum level does not arise due to differences in the pointing of
the telescope or the notorious HEB-band standing waves. We find the
uncertainty in the continuum level to be the most pronounced for the
$P(4)$ line at 1787~GHz.  We have evaluated the effect of these
variations and included them in our estimate of uncertainties of the
column densities (Sec. 3.1).

\begin{figure}
\begin{center}
\includegraphics[width=0.4\textwidth]{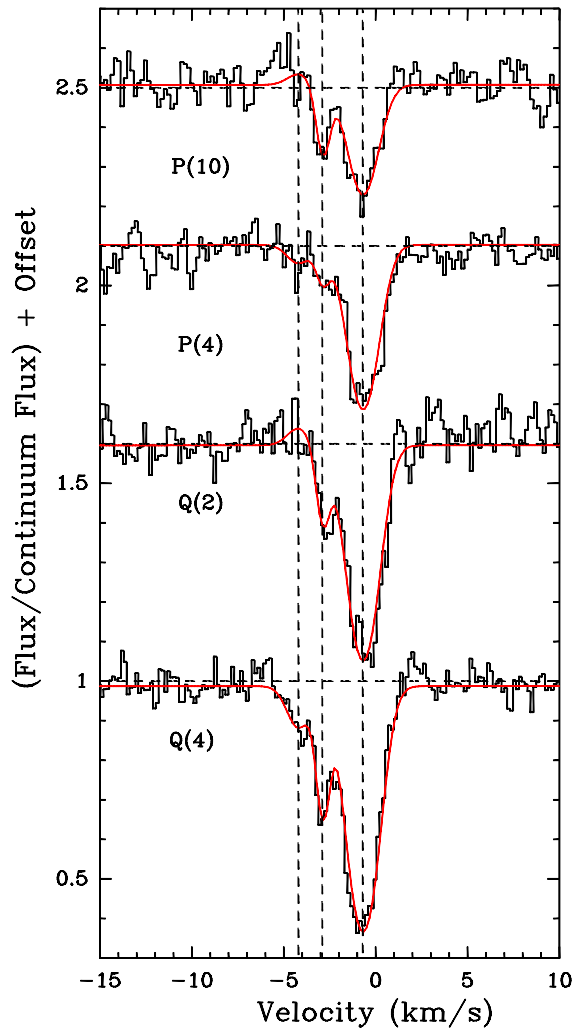}
\caption{ HIFI observations of the \cthree\ lines in DR21(OH). The
spectra are corrected for the DSB continuum and normalized to the SSB
continuum level.  The $Q(2)$, $P(4)$ and $P(10)$ spectra are shifted
consecutively by +0.5, so that the plots do not overlap. The red line
shows the simultaneous fit with 3 Gaussian velocity components (at
$-0.7$, $-2.9$ and $-4.2$~\kms). Note that the 3rd component is only
detected above the noise in $Q(4)$ and $P(4)$. The ``negative"
absorption bumps at this velocity fitted in $Q(2)$ and $P(10)$ is within
the noise.
\label{fig_c3spec}}
\end{center}
\end{figure}

Fig.~\ref{fig_c3spec} shows the observed spectra normalized to the
(single sideband) continuum level. The observed \cthree\ absorption
features occur around the systemic velocity ($\sim -1$~km/s) of an
infalling subfilament on DR21(OH) \citep[see][]{schneider2010} and hence
are associated with the source only and not with foreground material.
The PRISMAS observations are not sensitive enough to detect \cthree\
lines in absorption in the diffuse gas along the line of sight to
DR21(OH) \citep{mookerjea2010}.

As part of the PRISMAS program, in addition to the target lines detected in
absorption towards DR21(OH), a large number of emission lines falling
within the multiple observed bands were also detected. Here we use the two
spectral lines of CCH at 523.97 and 524.03~GHz in DR21(OH) in HIFI Band
1a.

Table~\ref{tab_specdata} summarizes the spectroscopic parameters of the
species and transitions observed with {\em Herschel}.

\begin{figure}
\begin{center}
\includegraphics[angle=-90,width=0.4\textwidth]{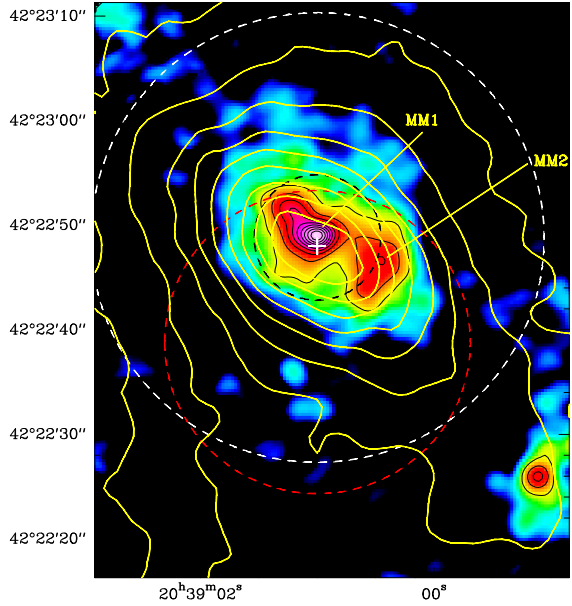}
\caption{The 3\,mm PdBI continuum image of DR21(OH) in color and black
contours overlaid with yellow contours of 450~\micron\ image
observed with JCMT SCUBA. Also overlaid on the continuum image are the
FWHM beam sizes for spectroscopic observations.  The dashed lines show
the HIFI (12\arcsec\ (black) \& 43\arcsec\ (white)) and IRAM 30m
(29\arcsec, red) beams.  The white cross denotes the center of the HIFI
beams.
\label{fig_overview}}
\end{center}
\end{figure}

\subsection{IRAM 30m data: CCH \& \cthreehtwo}

The ground state transition of CCH at 87.316 GHz (N=1-0,
J=3/2-1/2, with three hyperfine components) and the
2$_{1,2}$-1$_{0,1}$ transition of $c$-C$_3$H$_2$ at 85.338~GHz were
observed by \citet{gerin2011} using the IRAM-30m telescope in 2006,
August and December.  Table~\ref{tab_iramspec} presents both
spectroscopic and observational parameters of the lines observed with
IRAM 30m.  The primary aim of these observations was the study of CCH
and \cthreehtwo\ in the diffuse and translucent clouds along the line
of sight to DR21(OH).  However, as do the HIFI data, these
observations also detect the emission lines arising from the source
DR21(OH) itself. 

\subsection{Position of beams relative to the continuum sources in
DR21(OH)}

Figure~\ref{fig_overview} presents the 12\arcsec\ (\cthree) and
43\arcsec\ (CCH at 522~GHz) beams of HIFI and the 29\arcsec\ IRAM 30m
beams overlaid on the images of  3\,mm and 450~\micron\ continuum
emission from DR21(OH).  The 3~mm continuum emission (color \& black
contours) was observed with PdBI \citep{csengeri2011} and the
450\micron\ (yellow contour) continuum data were obtained from the JCMT
archives. We see that the 3~mm interferometric map (restored to a beam
of 2\arcsec) clearly resolves the continuum sources MM1 and MM2 and also
shows evidence for extended emission from the envelope. The 450~\micron\
JCMT data, at a resolution of 8\arcsec\ shows that the continuum
emission is primarily dominated by the surrounding extended envelope.
The \cthree\ beam of HIFI covers a significant part of the source MM2 as
well. The 29\arcsec\ IRAM 30m beam almost completely covers the larger
region sampled by the 522~GHz HIFI beam. Thus Fig.~\ref{fig_overview}
shows that the observing beams for all the spectral observations
presented in this paper cover regions including MM1, MM2 and the
surrounding envelope.  The spectral indices of the continuum emission
derived so far from the 3~mm (PdBI) and 1.4~mm \citep[SMA][]{zapata2012} data
are not consistent with thermal emission from dust. It is thus difficult
to characterize the contribution of MM1 and MM2 to the continuum
emission. However, at the beamsizes of our \cthree\ observations at THz
frequencies it is likely that the continuum is dominated by the emission
from the extended envelope and hence variation due to the contribution
of the individual sources is likely to be small.  We thus assume that all the
\cthree\ absorption is of the same continuum, produced primarily from
the extended envelope.

\section{Results}

\subsection{\cthree\ column densities}

The observed \cthree\ absorption spectra show evidence for multiple
velocity components (Fig.~\ref{fig_c3spec}). The $Q(4)$ transition
clearly shows three velocity components, while the other transitions
distinctly show two components.  In a global fitting scheme we have
assumed each \cthree\ transition to be having contributions from three
velocity components and then fitted all the observed transitions
simultaneously using multi-component gaussians.  The parameters being
fitted simultaneously in each transition are the common \vlsr, $\Delta
v$ for each of the three velocity components and the strength of the
absorption dips, separately for each line.  Thus, we have assumed all
three velocity  components to be present in all transitions even if the
components are not explicitly seen, since we do not attribute the
non-detection to any physical or chemical reason rather to the lower
strength of the transition relative to the achieved rms of our
observations.  All the four spectra show evidence for two major velocity
components centered at \vlsr\ = $-2.85\pm0.04$~\kms\ and \vlsr\ =
$-0.68\pm0.02$~\kms, with the latter component being stronger. The third
velocity component detected only in $Q(4)$ and marginally in the $P(4)$
spectra is at $-4.2\pm0.2$~\kms. We have checked that for a particular
transition in which the third component is undetected the results of
fitting do not change significantly for the well detected velocity
components by the assumption of the presence of the third component.
Table~\ref{tab_gaussfit} presents the single side-band continuum levels
(estimated from double side-band temperatures assuming a side-band gain
ratio of 1) and the results of the Gaussian fit along with their
uncertainties.

We compare the observed velocity components at $-0.7$, $-2.9$\,\kms\ and
$-4.2$\,\kms\ with the known velocities in the region.  The LSR
velocities of the sources MM1 and MM2 are $-4.1$ and $-0.7$~\kms,
respectively, and at a resolution of 12\arcsec\ the peak of C$^{18}$O
emission is observed to have a \vlsr\ of $-3.2$~\kms, which most likely
arises from the envelope surrounding MM1 and MM2
\citep{mangum1992,wilson1990}. Our observation is centered on MM1; MM2,
located 8\farcs3 away from MM1, lies within the HIFI beam. Thus, 
although we can not resolve MM1 and MM2 spatially,  the velocities of
the two sources and the envelope differ sufficiently from one another to
allow identification of the three components in the \cthree\ spectra. We
detect \cthree\ absorption due to MM2 and the envelope in all
transitions, while MM1 appears only in the stronger $Q(4)$ transition and
marginally in $P(4)$ transition.  Based on the $J=4$ column densities
derived for the $-4.2$~\kms\ component assuming $T_{\rm rot}$
between 50--70 K we estimate the expected absorption depths for the
$J=2$ level to be only 10\% and even less for the $J=10$ level and this
is comparable to the uncertainty in our measurements. Thus the
non-detection of the $-4.2$~\kms\ component in the $Q(2)$ and $P(10)$ is
consistent with the expected strengths of these lines.

\begin{table}[h]
\begin{center}
\caption{Single sideband continuum antenna temperatures (\Tc) and
parameters derived from simultaneous Gaussian fitting of all the line
profiles and the lower state \cthree\ column densities ($N_{\rm l}$)
estimated from the fitted intensities. The errorbars in $N_{\rm l}$
include errors due to both uncertainty in the continuum level and the
line fitting procedure.
The velocity components are as follows: Component 1:
$v_{\rm cen}$ = $-0.68\pm0.02$~\kms, $\Delta V$ = 1.85$\pm0.51$~\kms,
Component 2: $v_{\rm cen}$ = $-2.85\pm0.04$~\kms, $\Delta V$ =
0.86$\pm0.08$~\kms\ and Component 3: $v_{\rm cen}$ = $-4.2\pm0.2$~\kms,
$\Delta V$ = 1.4$\pm$0.4~\kms. Velocity component 3 is detected only above
the noise level in $P(4)$ and $Q(4)$.
\label{tab_gaussfit}}
\begin{tabular}{lccrr}
\hline
\hline
Transition &\Tc & $V_{\rm cen}$ & $\int{\tau d\rm v}$ & $N_{\rm l}$\\
& \hspace*{0.3cm}[K] & [km/s] &  [km/s] &  [10$^{14}$\cmsq] \\
\hline
$P(10)$& 7.1$\pm$0.4         & $-$0.7 & 0.60$\pm 0.07$ & 1.2$\pm$0.08 \\
       &                     & $-$2.9 & 0.17$\pm 0.04$& 0.3$\pm0.06$ \\
       &                     & $-$4.2 & \ldots & \ldots \\
&&&&\\
$P(4)$ & 7.0$\pm$1.3         & $-$0.7 & 1.00$\pm 0.06$ & 2.5$\pm0.1$  \\
       &                     & $-$2.9 & 0.08$\pm 0.03$& 0.2$\pm0.07$  \\
       &                     & $-$4.2 & 0.07$\pm 0.03$& 0.17$\pm0.08$ \\
&&&&\\
$Q(2)$ & 7.1$^{+0.6}_{-0.5}$ & $-$0.7 & 1.46$\pm 0.05$ & 1.2$\pm0.06$\\
       &                     & $-$2.9 & 0.19$\pm 0.03$& 0.2$\pm0.04$\\
       &                     & $-$4.2 & \ldots & \ldots \\
&&&&\\
$Q(4)$ & 6.7$^{+0.3}_{-0.7}$ & $-$0.7 & 1.79$\pm 0.04$ & 1.5$\pm0.06$ \\
       &                     & $-$2.9 & 0.33$\pm 0.03$ & 0.3$\pm0.03$ \\
       &                     & $-$4.2 & 0.16$\pm 0.03$ & 0.1$\pm0.06$ \\
\hline
Total $^{a}$ &               & $-$0.7 & \ldots &  9.1$\pm$2.3 \\
                     &       & $-$2.9 & \ldots &  2.2$\pm$0.4 \\
\hline
\end{tabular}
\end{center}

$^{a}$ Estimated using rotation diagram (Fig.~\ref{fig_c3rotdiag}).
\end{table}

\begin{figure}
\begin{center}
\includegraphics[width=0.5\textwidth]{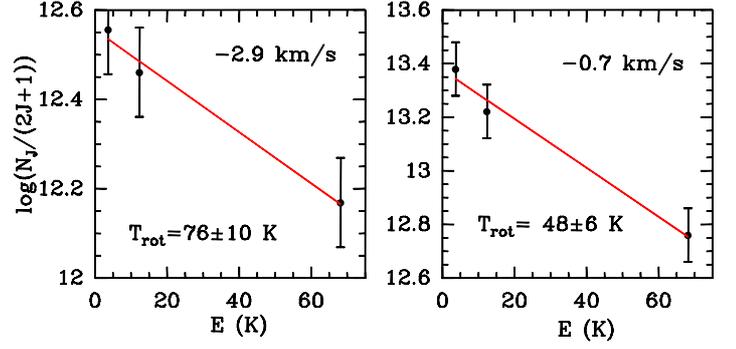}
\caption{Rotation diagram of \cthree\ for the velocity components at
$-0.7$~\kms\ and $-2.9$~\kms. The errorbars correspond to an estimated error
of 10\% in the state specific column densities and  are 
conservative estimates which include the uncertainties due to continuum
measurements as well as the noise in the spectra.
\label{fig_c3rotdiag}}
\end{center}
\end{figure}

Taking into consideration the uncertainties in the measurements
described above, we estimate the state-specific column densities to be
$\sim 10^{14}$~\cmsq\  and $\sim 10^{13}$~\cmsq\ for the $-0.7$~\kms\
and $-2.9$~\kms, components, respectively
(Table~\ref{tab_gaussfit}).  For $Q(4)$ \& $P(4)$ the additional
absorption feature at $-4.2$~\kms\ corresponds to a $J=4$ column density
of $\sim 10^{13}$~\cmsq.  Given the marginal detection at $P(4)$ and
non-detection at $Q(2)$ we propose the $J=4$ column density for the
$-4.2$~\kms\ component to be an upper limit. Using the $J$ = 2, 4 and
10 levels we construct rotation diagrams and obtain rotation
temperatures of 76$\pm20$~K and 48$\pm$6~K for the velocity components
at $-2.9$~\kms\ and $-0.7$~\kms\ respectively
(Fig.~\ref{fig_c3rotdiag}). The total \cthree\ column densities for the
$-2.9$~\kms\ and $-0.7$~\kms\ components are (2.2$\pm$0.4)$\times
10^{14}$~\cmsq\ and (9.1$\pm$2.3)$\times 10^{14}$~\cmsq, respectively.
Thus, the $-0.7$~\kms\ component corresponds to the colder and higher
column density gas.

\subsection{CCH column densities}

\begin{table}[h] 
\caption{Parameters derived from fitting the CCH (observed with HIFI 
and IRAM) and \cthreehtwo\ lines with LTE models using XCLASS.
Parameters without errorbars were held constant for the fitting.
\label{tab_xclass}}
{\small
\begin{tabular}{lccccc} 
\hline 
\hline 
Species & Size & T & $N$ & $\Delta v$ & $v_{\rm LSR}$ \\ 
& \arcsec & K & 10$^{14}$\cmsq & \kms & \kms \\ 
\hline
&&&&&\\ 
CCH-87     &  15  &  110$\pm$15  &  19.7$\pm$3.0   & 1.6$\pm$0.1  &  $-0.6\pm0.1$\\ 
           &  30  &   9$\pm2$   &   23.4$\pm$1.3   & 3.5$\pm$0.1  &  $-2.7\pm0.1$\\ 
           &  15  &  111$\pm$10  &  60.3$\pm$4.0   & 1.7$\pm$0.1  &  $-4.2\pm0.1$\\ 
&&&&&\\ 
CCH-524    &  30  &  33$\pm5$  & 1.2$\pm0.3$  &  5.5$\pm0.2$ &  $-2.8\pm 0.2$\\ 
           &  15  & 101$\pm11$ & 1.4$\pm0.3$  &  4.5$\pm0.2$ &  $-4.6\pm 0.2$\\ 
&&&&&\\ 
C$_3$H$_2$ & 30 & 31$\pm4$ & 0.7$\pm0.1$ & 3.3$\pm0.1$ & $-2.7\pm 0.1$\\
           & 15 & 78$\pm6$ & 0.9$\pm0.2$ & 1.5$\pm0.2$ & $-4.5\pm0.2$\\ 
&&&&&\\ 
\hline 
\end{tabular} 
}
\end{table}

\begin{figure} 
\begin{center}
\includegraphics[width=0.45\textwidth]{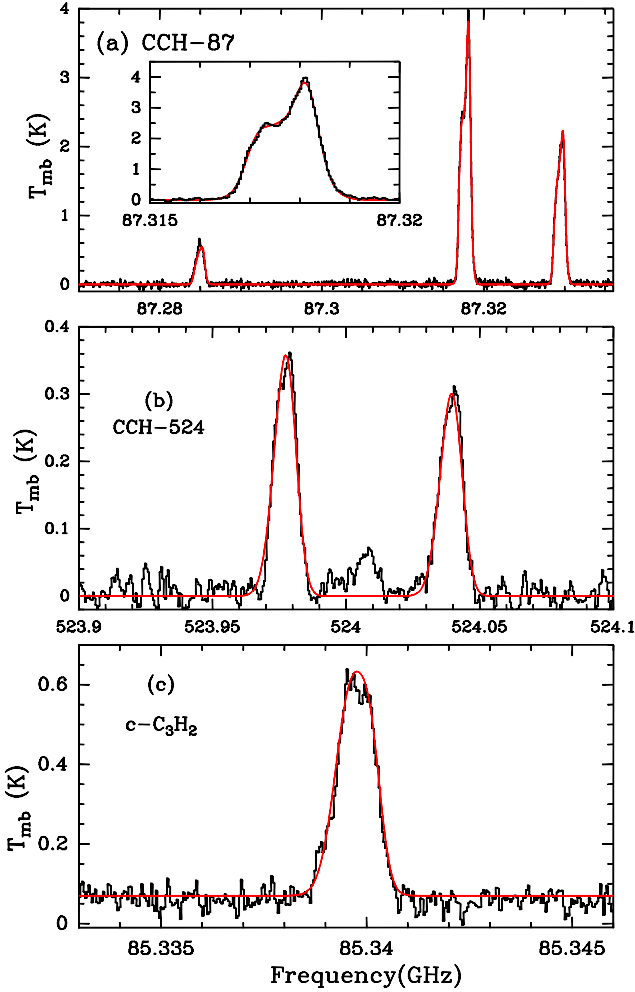}
\caption{Spectra of {\bf (a)} CCH at 87.316~GHz, {\bf (b)} CCH at 
523.9~GHz, and {\bf (c)} C$_3$H$_2$ at 85.338~GHz fitted with LTE 
models using XCLASS. (a) also shows the main hyperfine component of
87.316~GHz as inset to demonstrate the velocity structure better.
\label{fig_xclass} } 
\end{center} 
\end{figure}

In the PRISMAS observations we have detected 5 transitions of CCH at
frequencies around 87.3\,GHz and 524.0\,GHz. We have first
attempted to derive a self-consistent LTE model which explains  the
intensities of all the CCH transitions using a combination of XCLASS
\citep{schilke1999,comito2005} and MAGIX \citep{bernst2011}. XCLASS
generates synthetic spectra consisting of multiple velocity components
as well as hyperfine structure components for all transitions of
different chemical species.  XCLASS accesses the molecular databases
CDMS and JPL and models each molecule with the following free
parameters: source size, temperature, column density, line width and
velocity offset relative to the systemic velocity of the source, and
derives the column density corresponding to the different velocity
components detected in absorption and emission in the observed spectra.
The source size refers to the size of the cloud contributing to the
emission or absorption spectra. The synthetic spectra generated with
XCLASS were fitted to the observed spectra using MAGIX, an iterating
engine that allows automatic minimization to constrain model parameters.
We have assumed that the hyperfine lines have the same excitation
temperatures.  We note that the fitting procedure can involve
considerable degeneracy between the source size, the excitation
temperature and the column density.  In the absence of more accurate
information we have fixed the source sizes for MM1 and MM2 to 15\arcsec\
and the size for the envelope to 30\arcsec.  Using XCLASS/MAGIX  we
could not obtain an LTE model which consistently explained the
intensities of all the 5 lines of CCH. We found that the 87\,GHz lines
have intensities much higher than the expected values based on the
524\,GHz intensities  under the assumption of LTE. We have thus fitted
two different LTE models to the 87~GHz and 524~GHz spectra of CCH.
Figure~\ref{fig_xclass}(a) and (b) show the observed CCH spectra along
with the LTE models fitted using XCLASS and the fit parameters are
presented in Table~\ref{tab_xclass}.  The errorbars for the
parameters in Table~\ref{tab_xclass} represent the range of values over
which the best fit model does not change significantly.

From the intensities of the 87.316\,GHz CCH lines we estimate the
LTE temperatures of the $-0.5$, $-2.9$ and $-4.2$~\kms\ components
contributing to the CCH emission to be 110\,K,  9~K, and 111~K
respectively.  The CCH transition at 524~GHz shows evidence of only two
velocity components at $-2.8$~\kms\ and $-4.6$~\kms, with temperatures
of 33~K and 101~K respectively.  

While the temperatures of the envelope derived from the LTE models of
the 87~GHz and the 524~GHz transitions are not too discrepant, the total
CCH column densities of the envelope ($v_{\rm cen}=-2.7$~\kms) estimated
from the two transitions are 2.3$\times 10^{15}$~\cmsq\  and 2.9$\times
10^{14}$~\cmsq\ respectively, hence differing by a factor $\sim 8$.
This discrepancy in column densities indicates two possibilities: First,
it is likely that  the excitation of CCH is non-LTE so that, the 87\,GHz
transition with a low critical density is thermalized and traces the
total column density better, but a commensurate number of molecules in
the higher energy levels probed by the 524\,GHz lines are not present.
Secondly,  owing to the difference in the pointing centers of the
beams the 87~GHz data could also be tracing the emission from the
extended ridge in DR21(OH), whereas the higher-$J$ lines at 524\,GHz
primarily trace the DR21(OH) hot core and the envelope. Though
difficult to quantify, the difference in coupling of the emission with
the beams at 87 and 524 GHz, which are of different sizes and
differently centered could contribute to part of the discrepancy. 
For consistent comparison of the abundances of CCH, \cthreehtwo\ and
\cthree\ abundances in the envelope, we use the column densities
derived from the 87~GHz CCH transitions as an upper limit for chemical
models discussed later.

\subsection{LTE Modelling of \cthreehtwo}

We have also derived an LTE model for the single \cthreehtwo\ line using
a two-component Gaussian profile corresponding to the velocities around
$-2.7$~\kms\ and $-4.5$~\kms.  Fig.~\ref{fig_xclass}(c) shows the
observed \cthreehtwo\ spectrum fitted with a two-component model using
XCLASS.

\subsection{Temperature structure in DR21(OH)}

Based on the results of synthesis imaging of thermal NH$_3$ and
detection of the (7,7) transition of NH$_3$, which is 535~K above the
ground state, the kinetic temperature of the gas associated with MM1
(\vlsr = $-4.1$~\kms) is estimated to be $>$ 80~K
\citep{mangum1992,mauersberger1986}.  \citet{wilson1990} derived a
kinetic temperature of 34~K for this region using NH$_3$ observations at
a resolution of 40\arcsec.

We note that the 43\arcsec\ beamsize for the 524~GHz HIFI
observations and the 29\arcsec\ beamsize of IRAM 30~m
(Fig.~\ref{fig_overview}) include MM1, MM2 and the envelope.  Hence the
observed CCH emission is primarily due to the molecular gas surrounding
all these sources which is traced in CO and C$^{18}$O.  Using CCH and
\cthreehtwo\  we derive the temperature of the component corresponding
to the envelope to be 10--30~K and the temperature of the component
associated with MM1 to be 100--110~K. The  data on carbon chain molecules
are consistent with the presence of high temperatures in MM1 and MM2 and
agree with the previous determinations.

\section{Modeling the Observed Chemical Abundances in the Envelope}

\subsection{The OSU Gas Grain Models}

\begin{table}[h]
\begin{center}
\caption{Comparison of observed abundances of the chemical species in the envelope
of DR21(OH) and other sources.
\label{tab_obsabund}}
\begin{tabular}{lllll}
\hline
\hline
Species &  \multicolumn{2}{c}{DR21(OH)} & \multicolumn{1}{l}{Lukewarm} & \multicolumn{1}{l}{Hot} \\
 &  &  &  Corino$^{b}$ & Corino$^{c}$ \\
& $N$ & $X^{a}$ & $X$ & $X$\\
& cm$^{-2}$  &$\times 10^{-9}$  & $\times 10^{-9}$ & $\times 10^{-9}$ \\
\hline
\cthree\  & 2.2$\times 10^{14}$ & 0.63 & \ldots & \ldots\\
&&&&\\
CCH   & 2.3$\times 10^{15}$ & 6.6 & (5.3--12.8) & 0.05\\
(87 GHz)  &  &  & &\\
CCH   & 2.9$\times 10^{14}$ & 0.8 & (5.3--12.8) &0.05\\
(524 GHz)   &  &  & &\\
\cthreehtwo\ &  6.8$\times 10^{13}$ & 0.19 & 0.22 & \ldots$^{d}$\\
\hline
\end{tabular}
\end{center}

$^{a}$ Estimated using N(H$_2$) = 3.5$\times 10^{23}$~\cmsq\
(Table ~\ref{tab_srcprop}).\\
$^{b}$ \citet{sea09,sea09b}\\
$^{c}$ \citet{sea09,cea03,bisschop07}\\
$^{d}$ Observations not available, \citet{hhh11} model ($n_{\rm H_2}=5\times 10^5$ cm$^{-3}$) predicts X(\cthreehtwo)$\approx
10^{-9}$ in lukewarm corinos (T$\sim 30$~K) and X(\cthreehtwo)$\approx 10^{-7}$ in hot cores (T$\sim 100$~K).\\
\end{table}

We model the observed abundances of the three chemical species \cthree,
\cthreehtwo\ and CCH, (Table ~\ref{tab_obsabund}) in the warm envelope
around DR21(OH) using the Ohio State University (OSU) gas-grain code
with a warm-up \citep{hhl92,gh06}. Although slightly shifted in
velocity from the envelope at $-4.2$~\kms\ detected in \cthree, for
CCH-524 and \cthreehtwo\ we consider the $-4.5$~\kms\ velocity component
to correspond to the envelope. Since the models are primarily for dense
star forming regions and the low-frequency CCH transition (87\,GHz)
likely trace an additional extended component, we use the CCH abundance
derived from the IRAM data as an upper limit to compare with the outcome
of the model. \citet{motte2007} estimated the average volume density of
the entire region to be 10$^6$~\cmcub, while the MM1 and MM2 dense cores
reach densities larger than 10$^7$~\cmcub\ (Table~\ref{tab_srcprop}).
Thus for the envelope we presently consider models with constant gas
densities $n_{\rm H_2} =  5\times 10^{5}$, $
1\times10^{6}$,~and~$5\times 10^{6}$~cm$^{-3}$. 

The gas-grain network considers 7166 reactions involving a total of 668
gaseous and surface species, where the surface species are identified
with (s). The physical parameters and initial chemical abundances
adopted are the same as used by \citet{gwh07} and \citet[Tables 1 \&
2;][]{hhg08}.  In this approach, a one-point parcel of material
undergoes an initially cold period of $T_{0}=10$~K with a duration of
$t=10^5$ yr, followed by a gradual temperature increase.  A heating
timescale of $t_{\rm{h}}=0.2$~Myr is adopted following \citet{gh06},
where we use values for the maximum temperature $T_{\rm{max}}$ of 30 K
and 50 K, so that the model reaches $T_{\rm max}$ by $t\approx
0.3$\,Myr. A total of six combinations of $T_{\rm max}$ and $n_{\rm
H_2}$ are considered, as outlined in Table~\ref{tab_modpar1}.  In
contrast to more complex hydrodynamic approaches \citep{aik08}, here the
other physical parameters (Table~\ref{tab_modpar2}) are assumed to be
homogeneous at any given time and from one model to another. The
advantage of this approach is that it allows a more detailed look at the
chemical processes and the roles of individual reactions.  A similar
procedure was previously used to simulate the formation of hydrocarbon
chains in the envelopes of low mass protostellar envelopes by
\citet{hhg08} and \citet{hhh11}, where the latter model also considered
the linear C$_3$ detected toward W31C and W49N \citep{mookerjea2010}

The models include photodesorption rates for CO(s), N$_2$(s), H$_2$O(s),
and CO$_2$(s) based on the measurements of \citet{oberg1,oberg2,oberg3}.
The adopted processes and rates are species-dependent, rather than
dependent on a single parameter.

The production of hydrocarbon chains during the warm-up has been
attributed to a process beginning with grain surface chemistry.  The
elevated temperature allows methane to evaporate from the ice mantles
surrounding dust particles and act as a precursor for a carbon-chain
rich ion-molecule chemistry.  \citet[Fig.6;][]{hhh11} demonstrated that
the abundance of \cthree\ is subject to such a process.  In particular,
the \cthree\ is formed via gas-phase chemistry involving atomic C at
early times, and depletes sharply at $t \sim 10^{5}$ yr, after which
time most available atomic C is incorporated into CO.  In the absence of
grain surface chemistry, the \cthree\ will continue to deplete to
abundances below 10$^{-12}$ by 1 Myr.  In models including grain surface
chemistry, the abundance of \cthree\ is re-generated following the
release of CH$_4$(s) and subsequent gas phase chemistry during the
warm-up.  The abundances and reactions discussed in this section refer
to the gas phase.

\begin{table}[t]
\begin{center}
\caption{Parameters for the models explored for \cthree\ chemistry in
DR21(OH)}
\label{tab_modpar1}
\begin{tabular}{lcc}
\hline
\hline
Model & $T_{\rm max}$ (K) &$n_{\rm H_2}$ (cm$^{-3}$)\\
\hline
1 & 30 & $5\times 10^5$ \\
2 & 30 & $1\times 10^6 $  \\
3 & 30 & $5\times 10^6 $  \\
4 & 50 & $5\times 10^5 $  \\
5 & 50 & $1\times 10^6 $  \\
6 & 50 & $5\times 10^6 $  \\
\hline
\end{tabular}
\end{center}
\end{table}

\begin{table}[t]
\begin{center}
\caption{Parameters assumed to remain constant in present models.}
\label{tab_modpar2}
\begin{tabular}{lcll}
\hline
\hline
Parameter &Value  &Unit \\
\hline
$^b$ Cosmic Ray Ionization Rate $\zeta$ & $1.3\times 10^{-17}$& s$^{-1}$\\
$^b$ Visual Extinction $A_V$ & 10 &   \\
$^{a,b}$ Grain Radius $a_{\rm d}$ & $1\times 10^{-5}$ & cm \\
$^{a,b}$ Non-thermal Desorption Yield $a_{\rm RRK}$ & 0.01 &  \\
$^{a,b}$ Initial Temperature $T_{\rm i}$ & 10 & K\\
$^{a,b}$ Warm-up Timescale $t_{\rm sc}^{a,b}$ & $2\times 10^5$ & yr\\
$^b$ Warm-up Offset Time $t_{\rm off}$ & $1\times 10^5$ & yr \\
\hline
\end{tabular}
\end{center}
$^{a}$ \citet{gh06}; $^{b}$ \citet{hhg08}\\
\end{table}


\subsection{Model results}

\begin{figure*}
\begin{center}
\includegraphics[width=0.23\textwidth, angle=-90]{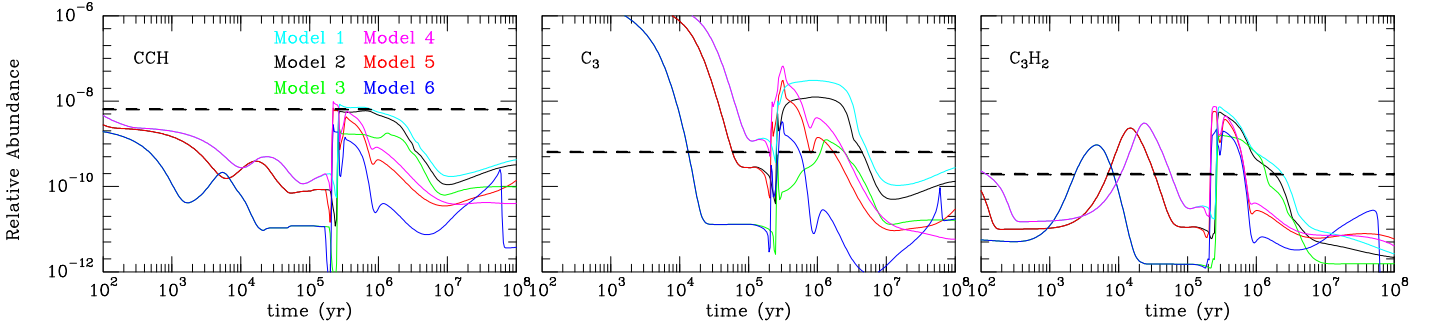}
\caption{Comparison of observed abundances in the envelope (shown by
dashed lines) with the results of chemical models involving warming up
of grains for  different combinations of hydrogen density ($n_{\rm
H_2}$) and final warm-up temperatures.  The model calculates abundances
relative to hydrogen, and the observed column densities of the different
species have been converted to fractional abundances assuming N(H$_2$) =
3.5$\times$ 10$^{23}$~\cmsq.  In all the models the warm-up occurs
between 0.1 and 0.3\,Myr starting from an initial temperature of 10~K,
similar to the temperature profiles that appear in
\citet[Figs.2 \& 3;][]{hhg08}.
\label{fig_chemmodel}}
\end{center}
\end{figure*}

The time evolution of fractional abundances is presented for \cthree,
CCH and \cthreehtwo\ in Fig.~\ref{fig_chemmodel}, along with the
corresponding observed values. In all the models the warm-up occurs
between 0.1 and 0.3\,Myr starting from an initial temperature of 10~K,
similar to the temperature profiles that appear in
\citet[Figs.2\&3;][]{hhg08}.  We observe the following general trends in
the abundance profiles shown in Fig.~\ref{fig_chemmodel}:   For the
lower temperature models (1-3), the molecular abundances attain values
comparable to the observed abundances for a much longer period of time
as compared to the higher temperature models (4-6). Further, for both
\cthree\ and \cthreehtwo\ in Models 4-6, the maximum abundance
decreases with increasing values of $n_{\rm H_2}$.  

Fig~\ref{fig_chemmodel} shows that all six models produce CCH to a peak
abundance of $X=0.2-1 \times 10^{-8}$ during the warm-up.  The abundance
depletes more rapidly for Models 4-6,
while the abundance remains elevated for a longer period for 
Models 1-3.  Moreover higher densities produce a slightly smaller
CCH abundance following the warm-up.  The \cthree\ abundance depletes
more rapidly for 
Models 4-6, and smaller
abundances are produced for larger density during this period.  All the
models produce a similar abundance of \cthreehtwo\, in the range of
$X=0.5-1 \times 10^{-8}$, following the warm-up.  The temperature and
density effects seen in the previous two species also appear for
\cthreehtwo\, but to a lesser extent.  

The primary reason for the difference in abundance of the chemical
species considered here, for  $T_{\rm max}=30$~K and 50 K models is that
the higher temperature will release additional species from the icy
grain mantles to the gas.  This leads to additional, competitive
reaction pathways with the overall effect of a comparative depletion of
abundance of these particular species.   A similar effect is visible at
the onset of the warm-up, when the abundances of the three species
deplete sharply just prior to the steep increase attributed to the
warm-up.

\subsection{Major Chemical Reactions}

Table~\ref{tbl-rxns} summarizes the most important chemical reactions
occurring in the gas-phase during warm-up  as determined by the
gas-grain chemistry model \citep{gh06}.  We emphasize that the species
produced here also result from other ion-molecule reactions, and that
Table~\ref{tbl-rxns} lists only the most direct and salient mechanisms.
The liberated CH$_4$ reacts with  C$^+$ to form the intermediate
species C$_2$H$_{2}^{+}$ and C$_2$H$_{3}^{+}$ via reactions (1) and
(3).  Subsequently reaction of H$_2$ with C$_2$H$_{2}^{+}$ in (2)
produces C$_2$H$_{4}^{+}$.   The dissociative recombination reaction
of an electron with these intermediate ions then produces either
C$_2$H$_2$ by reactions (2) or (5).  Further reaction of CH$_4$
with the intermediate ions can produce C$_3$H$_5^+$ via reactions (6)
and (7), while the dissociative recombination of C$_3$H$_5^+$ produces
C$_3$H$_3$ via reaction (8).   Reactions of C and C$^+$ with C$_2$H$_2$
(9,11) produce C$_3$H and C$_3$H$^+$.  Further reaction of these
products (12,13) produce C$_3$H$_2^+$, and the reaction of CH$_4$ with 
C$_3$H$_2^+$ (14) produces C$_3$H$_3^+$.  The C$^+$ and C result from
partial desorption of CO(s) and subsequent reaction with He$^+$ during
the warm-up. These represent the major precursors for the observed
species, and provide the basis to examine each species individually.

For the chemical species considered here we identify the primary
formation reactions as follows: CCH is primarily formed by the
dissociative recombination of C$_2$H$_3^+$ in reaction (17).  The
evolution of \cthree\ proceeds in a somewhat different manner.
Ion-molecule chemistry forms a relatively large abundance of this
species at early times, followed by a depletion to grain surfaces.  This
depletion happens at slightly earlier times for models with larger
values of density.  Although \cthree\ is frozen on the dust grains,
during the warm-up, \cthree\ is not liberated directly from the grains,
but rather is produced in a reaction sequence starting with CH$_4$, 
whilst the \cthree (s) is more likely to participate in surface
reactions than to desorb.  The reaction of C + C$_2$H$_2$ produces
\cthree\ through reaction (19).  This mechanism and the secondary
pathways result in the formation of a peak in the abundance during and
after the warm-up period.  The formation of \cthreehtwo\ can proceed via
three pathways. The first mechanism is reaction (21), in which
C$_3$H$_3$ reacts with H.   The dissociative recombination of
C$_3$H$_3^+$ (22) is a second mechanism. Alternatively, reaction (23)
of CH with C$_2$H$_2$ is another important formation pathway.

\begin{table}[h]
\caption{Dominant reactions of observed and intermediate species
  following CH$_4$ sublimation.  The observed species are identified
  in bold type.             
\label{tbl-rxns}}
\renewcommand{\footnoterule}{}  
\begin{tabular}{llllllll}
\hline\hline       
& && C$_2$H$_2$: \\
\hline 
1.) &C$^+$ &+& CH$_4$ & $\rightarrow$ & C$_2$H$_3^+$ &+& H\\
2.)&C$_2$H$_3^+$ &+& e & $\rightarrow$ &C$_2$H$_2$ &+&  H  \\
3.) &C$^+$ &+& CH$_4$ & $\rightarrow$ & C$_2$H$_2^+$ &+& H$_2$ \\
4.)&C$_2$H$_2^+$ &+& H$_2$ & $\rightarrow$ &C$_2$H$_4^+$ \\
5.)&C$_2$H$_4^+$ &+ &e & $\rightarrow$ &C$_2$H$_2$ &+& 2 H  \\
\hline       
& &&  C$_3$H$_3$: \\
\hline 
6.) & C$_2$H$_3^+$ &+& CH$_4$ & $\rightarrow$& C$_3$H$_5^+$ &+& H$_2$ \\
7.)  & C$_2$H$_2^+$ &+& CH$_4$ & $\rightarrow$ &C$_3$H$_5^+$ &+& H \\
8.)  & C$_3$H$_5^+$ &+& e & $\rightarrow$ &C$_3$H$_3$ &+& H$_2$\\
\hline
& &&  C$_3$H$_3^+$: \\
\hline 
9.)     & C & + & C$_2$H$_2$ &$\rightarrow$& C$_3$H & + & H \\
10.)     & HCO$^+$ & +& C$_3$ &$\rightarrow$& C$_3$H$^+$ &+& CO \\
11.)      & C$^+$ & + & C$_2$H$_2$ &$\rightarrow$& C$_3$H$^+$ & + & H \\
12.)     & HCO$^+$ & +& C$_3$H &$\rightarrow$& C$_3$H$_2^+$ &+& CO \\
13.)     & C$_3$H$^+$ &+& H$_2$ &$\rightarrow$& C$_3$H$_2^+$ &+& H \\
14.) & C$_3$H$_2^+$ &+& CH$_4$ & $\rightarrow$& C$_3$H$_3^+$ &+& CH$_3$ \\
15.)  & HCO$^+$ &+& \cthreehtwo & $\rightarrow$ &C$_3$H$_3^+$ &+& CO \\
16.)    & C$_3$H$^+$ &+& H$_2$ &$\rightarrow$& C$_3$H$_3^+$ &\\
\hline
& && CCH: \\
\hline
17.)  & C$_2$H$_3^+$ &+& e & $\rightarrow$ & \bf CCH &+& 2 H \\
18.)  & HCO$^+$&+& \bf CCH &  $\rightarrow$ & C$_2$H$_2^+$ &+&  CO \\
\hline
& && C$_3$:\\
\hline
19.) & C &+& C$_2$H$_2$ & $\rightarrow$ & \bf C$_3$ &+&  H$_2$ \\ 
20.)& \bf CCH &+& \bf C$_3$ &  $\rightarrow$ & C$_5$ &+&  H \\ 
\hline
& && \cthreehtwo:\\
\hline
21.) & H &+& C$_3$H$_3$ & $\rightarrow $ &\bf \cthreehtwo\ &+& H$_2$\\
22.)        & C$_3$H$_3^+$&+& e &$\rightarrow $ & \bf \cthreehtwo\ &+& H \\
23.) & CH &+& C$_2$H$_2$ & $\rightarrow $ &\bf \cthreehtwo\ &+& H \\
24.) & HCO$^+$ &+& \bf \cthreehtwo\ & $ \rightarrow $ &C$_3$H$_3^+$ &+& CO \\
\hline                  
\end{tabular}
\end{table}

\subsection{Best fit chemical model }

In order to compare model results with observations on a more
quantitative basis, we compute the mean confidence level \citep{gwh07}
during the warm-up period.  In this quasi-statistical analysis
introduced by  \citet{gwh07}, the confidence level, $\kappa_i$, for
agreement between computed and observed abundances is defined as 
\begin{equation}
\label{eqn-kapdef}
\kappa_i = erfc \left( {|\log(X_i)-\log(X_{obs,i})|} \over { \sqrt(2) \sigma} \right)
\end{equation}

The index $i$ refers to individual chemical species under
consideration. The mean of the individual values of $\kappa_i$ thus
computed is denoted by $\kappa_{avg}$.  We consider the computed
abundance to ``fit" the observation if the abundance agrees within an
order of magnitude or better, and thus assume $\sigma=1$ so that one
standard deviation corresponds to agreement of one order of magnitude.
The related confidence level for such a factor of 10 agreement is
$\kappa_i=0.317$, and the value increases up to $\kappa_i=1$ for closer
agreement.  The accuracy of this analysis and the robustness of
$\kappa_{\rm avg}$ against individual values of $\kappa_i$ improves with
increasing size of the considered dataset. Since we have considered only
three chemical species, we use the $\kappa$-based analysis primarily as
a guideline to identify models which best reproduce the data.  The best
fit corresponds to the maximum average confidence level.

\begin{figure}
\begin{center}
\hspace*{-0.5cm}
\includegraphics[width=0.28\textwidth,angle=-90]{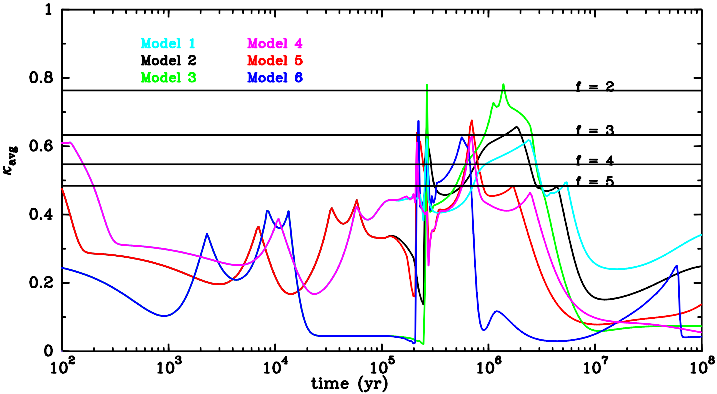}
\caption{ Mean confidence level $\kappa_{\rm avg}$ calculated for the chemical models as
explained in the text \citep{gwh07}.
\label{fig_kappa}}
\end{center}
\end{figure}

Fig.~\ref{fig_kappa} shows the $\kappa_{\rm avg}$ values for each model
that we have considered here, along with horizontal lines indicating
agreement within a factor ($f$) of 2, 3, 4, and 5.  The primary
interpretation of this diagram is that all six models reasonably fit the
observed abundances during the period following the warm-up.    
Models 1-3 fit the observation set within a factor of
2-4 or better for the time period of $t\sim 0.7-3$\ Myr for all three
density values.  
Models 4-6 fit the observation
set within a factor of 3-5 or better for the period of  $t\sim
0.6-0.8$\,Myr for 
Models 4 and 5 and for the earlier period  $t\sim
0.3-0.7$\,Myr  for Model 6.
Model 3 gives the best agreement with observations in terms of the value
and duration of the $\kappa_{\rm avg}$ function, while 
Model 2 is the second-best at meeting
these criteria. This analysis quantitatively asserts the level of model
agreement described in Fig.~\ref{fig_chemmodel}, and broadly shows
that the best sustained agreement of the models occurs in the time
following the warm-up rather than earlier or later.

\section{Discussion}

\subsection{Lower gas phase abundance of \cthree\ in MM1}

The angular resolution of the spectroscopic observations used in this
paper lies between 12 and 40\arcsec, so that MM1 and MM2 are not
resolved spatially. However, due to their markedly different velocities,
MM1, MM2 and the envelope are easily discernible in the spectra.
Interestingly, all \cthree\ spectra show primarily two velocity
components associated with MM2 and the envelope, with absorption due to
MM1 being detected only in $Q(4)$. In contrast, the chemically related
species CCH and \cthreehtwo\  that we have considered here show
components corresponding to only MM1 and the envelope.  Thus there is an
intrinsic difference in the distribution of \cthree\ compared to the
distribution of the other species. The abundance of \cthree\ in MM2
(1.2$\times 10^{-9}$) is only twice that in the  envelope (6.3$\times
10^{-10}$).  Based on Table~\ref{tab_srcprop} the total column density
of MM1 is $\sim 60\%$ of the column density of MM2. This suggests that
if the relative abundances of \cthree\ in MM1 and MM2  were similar, the
total column density of \cthree\ in MM1  would have been 2.5 times the
column density in the envelope so that the absorption dips due to MM1
would be similar to or stronger than the dip due to the envelope.  It
may be argued that the continuum backgrounds for the two components MM1
and the envelope may be different. However the continuum from MM1 is
expected to be stronger than the continuum from the envelope.  All of
the above arguments indicate that \cthree\ is preferentially depleted or
destroyed in MM1.


One major difference between MM1 and MM2 lies in the fact that MM1 has
already shown evidence for at least one high temperature hot core,
whereas MM2, though more massive shows no such components.  The chemical
network includes among others the reaction \cthree\ + H$_2$
$\rightarrow$ C$_3$H + H \citep{PFD88}, which becomes a primary
destruction pathway for \cthree\ when the temperature exceeds $\approx$
80 K.  As a result, models predict that \cthree\ will be depleted in
favor of \cthreehtwo\ in hot core conditions, as C$_3$H is also
destroyed by reaction with H$_2$. Additionally, the chemical models
presented here also show that the \cthree\ abundance is reduced at
larger densities.  MM1 is definitely at a higher density than the
envelope (Table~\ref{tab_srcprop}), so that the reduced abundance of
\cthree\ in MM1 relative to MM2 could also indicate that MM2 has a
density lower than MM1. We note that this contradicts the densities
derived from dust column densities (Table~\ref{tab_srcprop}), which is
not completely unexpected since the densities derived from the column
densities are beam averaged and need not always reflect the local
densities in a region.


However, an alternate possibility arises from the fact that the source
intrinsic continuum opacity can re-fill the absorption to a significant
extent, so that the column densities derived here are all primarily
lower limits \citep{mookerjea2010}.  It can thus be envisaged that in
the case of MM1, which has a stronger continuum emission, it is a
combination of the geometry of the source and the re-filling of the
absorption dip which reduces the ``visibility" of \cthree. A proper
evaluation of the effect  of the continuum opacity on the absorption
depth is possible only by constructing a complete radiative transfer
model of the entire region with accurate temperature and density
profiles, which are not yet available. 

Finally, since MM2 is estimated to be younger than MM1 and \cthreehtwo,
and CCH emission from MM2 are not detected, the possibility that  the
observed \cthree\ abundance in MM2 is due to the pre-warm-up gas-phase
chemistry can also not be ruled out.

\subsection{Comparison with \cthree\ in diffuse clouds} 

\cthree\ has been found to have an almost one-to-one correlation with
C$_2$ in diffuse clouds \citep{roueff2002}. Further, the abundance of
\cthree\ in diffuse clouds, (3--6)$\times 10^{-9}$ \citep{roueff2002},
is ten times larger than the abundance of \cthree\ derived in the dense
star forming environment of DR21(OH).  \citet{oka2003} had explained
the observed correlation between C$_2$ and C$_3$ in diffuse clouds, in
terms of a direct pathway of formation of \cthree\ from C$_2$ (see their
Fig. 4).  In the previous section, we examined the formation of C$_3$
during a warm-up period, and noted a relatively larger abundance of this
species prior to $t=10^5$~yr, the so-called ``early time".  We now
explore the chemistry of \cthree\ in diffuse regions and during the
early time period in dense gas regions, to examine the differences (if any).

The first step of the transformation of C$_2$ to C$_3$ in diffuse
clouds is the photoionization of C$_2$ to C$_2^+$ \citep{oka2003}. In
contrast, the models for the dense star forming regions assume darker
($A_V = 10$) conditions, so photoionization is not an effective
process. 

In dense models, the process begins with the radiative association of
C$^+$ + H$_2$ $\rightarrow$ CH$_2^+$. Next, CH$_2^+$ reacts with H$_2$
to form CH$_3^+$.  This is followed by the dissociative recombination of
CH$_3^+$ and CH$_2^+$ as competitive reactions to form CH, and then an
ion-neutral  reaction with C$^+$ to form C$_2^+$.  As shown in Fig.~4 of
\citep{oka2003}, the pathway then follows along the upper (C$_2$H)
branch to form C$_3$.  The C$_3$ is destroyed by processes that form
C$_4$, C$_4^+$, and C$_5$, which react further to primarily form C$_3$
in a cyclic process.  The abundance of C$_3$ grows until C$^+$ is
depleted at $t\sim 10$--100~yr, depending on the density, after which it
reaches a plateau until the C$_3$ is accreted onto grain surfaces.
Dissociative recombination reactions with C$_2$H$^+$, C$_2$H$_2^+$, and
C$_3^+$ along the pathway form C$_2$. The C$_2$ then reacts with O to
form CO, and as a result, the abundance falls to $X(\rm C_2) \sim
10^{-9}$, far below the $X(\rm C_2): X(\rm C_3)=40:1$ measured by
\citet{oka2003} for diffuse clouds.

Thus, C$_2$ is not the starting point for C$_3$ formation in dense cloud
models at early times, and also does not undergo a cycling process like
that of C$_3$.  As a result, the correlation between C$_2$ and C$_3$
abundance is not predicted by the dark cloud models. Rather the C$_2$
abundance evolution more closely follows the evolution of CCH from early
times until the warm-up begins than that of C$_3$.  This comparison
indicates that the chemistry of \cthree\ formation in the envelope of
DR21(OH) is definitely not the same as in diffuse clouds. Further, the
early time abundance of C$_3$ for dense regions exceeds the observed
abundance by three orders of magnitude, effectively excluding this
formation route as well.

\subsection{Warm Carbon Chain Chemistry in the envelope}

Comparison of the present model results and observations indicates
that the observed composition can be simulated with ion-molecule
chemistry following a moderate warm-up to 30-50~K. This also means
that \cthree\ formation in star forming regions can be explained in
terms of the moderate temperature (30~K) gas-phase chemical reactions
starting from CH$_4$ evaporated from the grain mantle only, and does
not require photochemistry of PAH molecules.

The observed abundances of C$_3$ are consistent with formation during a
warm-up model, or ``Warm Carbon Chain Chemistry'' (WCCC) \citep{sea07},
and related to the abundances of CCH, \cthreehtwo.
This suggests that the envelope conditions are more like those of
``lukewarm corinos'' surrounding low mass protostars than those of hot
cores. In other words the \cthree\ is not residing in a hot core.  The
abundances of the additional species chemically related to \cthree,
which we consider for the case of DR21(OH) provided further insight
regarding the envelope conditions.

Table~\ref{tab_obsabund} summarizes the observed abundances of the
different species in DR21(OH) as well as those available in the
literature for two types of sources: the lukewarm corinos  L1527 and
B228 and the hot corino IRAS 16293-2422.  We find that while \cthree\
has not been observed in any of the other sources, the abundances of the
other three species in DR21(OH) are closer to the abundances
observed in lukewarm corinos. Further, the chemical model that best
represents the observed abundances corresponds to a $T_{\rm max}$ of
30--50~K, a temperature that is much lower than the values expected in hot
cores/corinos. This suggests that the envelope of DR21(OH) is chemically
closer to a WCCC region such as a lukewarm corino envelope than to a hot
core. Further, the pre-depletion \cthree\ abundance (at $t\leq 10^4$~yr)
exceeds the observed values by a large factor. This implies that the
\cthree\ abundance is primarily maintained via ion-molecule chemistry in
the gas phase after CH$_4$ is desorbed from the surface of the dust
grains, and does not have a one-to-one correlation with C$_2$ as found in
diffuse clouds.  

\subsection{Chemical \& Dynamical Ages of the Region}

The chemistry of the envelope surrounding the embedded sources MM1 and
MM2 in DR21(OH) appears to have occurred over a period between 
(0.7--3)~Myr  including a cold period, warm-up, and extended time at $T
= 30$ or 50 K.  This time period appears to be on the longer side of the
evolutionary timescales of massive star forming cores. As detailed by
\citet{csengeri2011} the life time of the gas in  massive dense cores
is determined by the crossing times or the local free-fall times for the
molecular dense cores which is (5--7)$\times 10^4$~yr.  However the
large scale flows associated with Cygnus X (and DR21 in particular) are
massive enough to continuously replenish the mass of these cores
allowing them to remain ``active" for a much longer period of time and
the dynamical time of the most massive sub-filament in DR21 is $\sim
2$~Myr \citep{schneider2010}. Thus the chemical age of $\sim 1$~Myr
is consistent with the dynamical age of the region.

\section{Summary}

In the DR21(OH) region, using the high velocity resolution of HIFI, we
detect absorption due to \cthree\ from the envelope and the less evolved
core (MM2), and a rather weak signal associated to the most massive and
hottest region (MM1).  The abundance of \cthree\ in the envelope of the
hot core associated with DR21(OH) has been consistently explained along
with those of other species formed in the same chemical network using a
chemical model involving the warm-up of grains. The formation mechanism
of \cthree\ required to explain the observed abundances in dense star
forming regions is the gas-phase reaction of CH$_4$ desorbed from the
grain surfaces that have been warmed up.  It is not formed in a cycle
with C$_2$ as the starting point nor is it produced by the
photodestruction cascades of PAH molecules. For the envelope of DR21(OH)
the chemical models which best explain the abundances of \cthree\ and
other chemically related species correspond to Models 2 and 3, n$_{\rm H_2}$
=1--5$\times10^6$~\cmcub\ for $T_{\rm max}$ = 30~K , for a time period
between 0.7-3 Myr and  Model 6, n$_{\rm H_2}$ =$5\times10^6$~\cmcub\ for $T_{\rm
max}$ = 50~K, for a time period between 0.6-0.8 Myr.  The upper limit of
the timescale, though much larger than the typical lifetimes of hot
(warm) cores, is not unrealistic in the case of DR21(OH) where the
dynamic age of the most massive filament has been estimated to be 2 Myr. 

{}


\begin{thebibliography}{}

\bibitem[{\'A}d{\'a}mkovics et al.(2003)]{adamkovics2003}
{\'A}d{\'a}mkovics, M., Blake, G.~A., \& McCall, B.~J.\ 2003, \apj,
595, 235 

\bibitem[Aikawa et al.(2008)]{aik08} Aikawa, Y., Wakelam, V., Garrod,
R.T., \& Herbst, E. 2008, \apj, 674, 984

\bibitem[Araya et al.(2009)]{araya2009} Araya, E.~D., Kurtz, S., 
Hofner, P., \& Linz, H.\ 2009, \apj, 698, 1321 

\bibitem[Argon et al.(2000)]{argon2000} Argon, A.~L., Reid, 
M.~J., \& Menten, K.~M.\ 2000, \apjs, 129, 159 

\bibitem[Bernst et al.(2011)]{bernst2011} Bernst, I., Schilke, P.,
Moeller, T., et al.\ 2011, Astronomical Data Analysis Software and
Systems XX, 442, 505

\bibitem[Beuther et al.(2008)]{beuther2008} Beuther, H., Semenov, 
D., Henning, T., \& Linz, H.\ 2008, \apjl, 675, L33

\bibitem[Bisschop et al.(2007)]{bisschop07} Bisschop, S.~E.,
J{\o}rgensen, J.~K., van Dishoeck, E.~F., \& de Wachter, E.~B.~M.\
2007, \aap, 465, 913 

\bibitem[Cazaux et al.(2003)]{cea03} Cazaux, S., Tielens, A. G. G. M.,
Ceccarelli, C., Castets, A., Wakelam, V., Caux, E., Parise, B., \&
Teyssier, D. 2003, \apj, 593, L51

\bibitem[Cernicharo et al.(2000)]{cernicharo2000} Cernicharo, J.,
Goicoechea, J.~R., \& Caux, E.\ 2000, \apjl, 534, L199

\bibitem[Chandler et al.(1993)]{chandler1993} Chandler, C.~J., 
Moore, T.~J.~T., Mountain, C.~M., \& Yamashita, T.\ 1993, \mnras, 261, 694 

\bibitem[Comito et al.(2005)]{comito2005} Comito, C., Schilke, P.,
Phillips, T.~G., et al.\ 2005, \apjs, 156, 127

\bibitem[Csengeri et al.(2011)]{csengeri2011} Csengeri, T., 
Bontemps, S., Schneider, N., et al.\ 2011, \apjl, 740, L5 

\bibitem[Fish et al.(2005)]{fish2005} Fish, V.~L., Reid, M.~J., 
Argon, A.~L., \& Zheng, X.-W.\ 2005, \apjs, 160, 220 

\bibitem[de Graauw et al.(2010)]{deGraauw2010} de Graauw, T., et al.\
2010, \aap, 518, L6 

\bibitem[Galazutdinov et al.(2002)]{galazutdinov2002} Galazutdinov,
G., P{\v e}tlewski, A., Musaev, F., et al.\ 2002, \aap, 395, 969 

\bibitem[Garrod \& Herbst(2006)]{gh06} Garrod, R.T., \& Herbst, E.
2006, \aap, 457, 927

\bibitem[Garrod et al.(2007)]{gwh07} Garrod, R.T., Wakelam, V., \&
Herbst, E. 2007, \aap, 467, 1103

\bibitem[Gendriesch et al.(2003)]{gendriesch2003} Gendriesch, R, Pehl,
T.,~F., Winnewisser, G. et al.\ 2003, Zeit. Naturforsch., 58a,
129

\bibitem[Gerin et al.(2010)]{gerin2010} Gerin, M., De Luca, M., Black,
J.  et al.\ 2010, \aap, Herschel Special issue (in press)

\bibitem[Gerin et al.(2011)]{gerin2011} Gerin, M., Ka{\'z}mierczak,
M., Jastrzebska, M., Falgarone, E., Hily-Blant, P., Godard, B., \& de
Luca, M.\ 2011, \aap, 525, A116 


\bibitem[Giesen et al.(2001)]{giesen2001} Giesen, T.~F., Van
Orden, A.~O., Cruzan, J. D. et al.\ 2001, \apjl, 551, L181

\bibitem[Haffner \& Meyer(1995)]{haffner1995} Haffner, L.~M., \&
Meyer, D.~M.\ 1995, \apj, 453, 450

\bibitem[Hasegawa et al.(1992)]{hhl92} Hasegawa, T.I., Herbst, E., \&
Leung, C.M. 1992, \apjs, 82, 167

\bibitem[Hassel et al.(2008)]{hhg08} Hassel, G.E., Herbst, E., \&
Garrod, R.T. 2008, \apj, 681, 1385

\bibitem[Hassel et al.(2011)]{hhh11} Hassel, G.E., Harada, N., \&
Herbst, E. 2011, \apj, 743, 182

\bibitem[Hinkle et al.(1988)]{hinkle1988} Hinkle, K.~W., Keady,
J.~J., \& Bernath, P.~F.\ 1988, Science, 241, 1319

\bibitem[Maier et al.(2001)]{maier2001} Maier, J.~P., Lakin,
N.~M., Walker, G.~A.~H., et al.\ 2001, \apj, 553, 267

\bibitem[Mangum et al.(1991)]{mangum1991} Mangum, J.~G., Wootten, 
A., \& Mundy, L.~G.\ 1991, \apj, 378, 576 

\bibitem[Mangum et al.(1992)]{mangum1992} Mangum, J.~G., Wootten, 
A., \& Mundy, L.~G.\ 1992, \apj, 388, 467 

\bibitem[Mauersberger et al.(1986)]{mauersberger1986} Mauersberger,
R., Henkel, C., Wilson, T.~L., \& Walmsley, C.~M.\ 1986, \aap, 162,
199 

\bibitem[Mookerjea et al.(2010)]{mookerjea2010} Mookerjea, B., et al.\
2010, \aap, 521, L13 

\bibitem[Motte et al.(2007)]{motte2007} Motte, F., Bontemps, S.,
Schilke, P., et al.\ 2007, \aap, 476, 1243 

\bibitem[M{\"u}ller et al.(2005)]{mueller2005} M{\"u}ller, H.~S.~P.,
Schl{\"o}der, F., Stutzki, J., \& Winnewisser, G.\ 2005, Journal of
Molecular Structure, 742, 215 

\bibitem[{\"O}berg et al.(2007)]{oberg1} {\"O}berg, K.I., Fuchs, G.W., Awad, Z., et al. 2007, \apj, 662, L23

\bibitem[{\"O}berg et al.(2009a)]{oberg2} {\"O}berg, K.I., Linnartz, H., Visser, R., \&  van Dishoeck, E.F. 2009a, \apj, 693, 1209

\bibitem[{\"O}berg et al.(2009b)]{oberg3} {\"O}berg, K.I., van Dishoeck, E.F., \&  Linnartz, H., 2009b, \aap, 496, 281

\bibitem[Oka et al.(2003)]{oka2003} Oka, T., Thorburn, J.~A.,
McCall, B.~J., et al.\ 2003, \apj, 582, 823

\bibitem[Ott et al.(2010)]{ott2010}Ott S., 2010, in ASP conference
series “Astronomical Data analysis Software and Systems XIX”, Y.
Mizumoto, K. I. Morita and M. Ohishi eds., in press

\bibitem[Pety et al.(2005)]{pety2005} Pety, J., Teyssier, D.,
Foss{\'e}, D., Gerin, M., et al.\ 2005, \aap, 435, 885

\bibitem[Pilbratt et al.(2010)]{pilbratt2010} Pilbratt, G.~L.,
Riedinger, J.~R., Passvogel, T., et al.\ 2010, \aap, 518, L1 

\bibitem[Pineau des For\^{e}ts et al.(1988)]{PFD88} Pineau des
For\^{e}ts, G., Flower, D.R., \& Dalgarno, A. 1988, MNRAS, 235, 621

\bibitem[Plambeck \& Menten(1990)]{plambeck1990} Plambeck, R.~L., \&
Menten, K.~M.\ 1990, \apj, 364, 555 

\bibitem[Radi et al.(1988)]{radi1988} Radi, P.~P., Bunn, T.~L., 
Kemper, P.~R., Molchan, M.~E., \& Bowers, M.~T.\ 1988, \jcp, 88, 2809 


\bibitem[Reipurth \& Schneider(2008)]{reipurth2008} Reipurth, B., \&
Schneider, N.\ 2008, Handbook of Star Forming Regions, Volume I, 36 

\bibitem[Richardson et al.(1994)]{richardson1994} Richardson, K.~J.,
Sandell, G., Cunningham, C.~T., \& Davies, S.~R.\ 1994, \aap, 286, 555 

\bibitem[Roelfsema et al.(2012)]{roelfsema2012} Roelfsema, P.~R.,
Helmich, F.~P., Teyssier, D., et al.\ 2012, \aap, 537, A17

\bibitem[Roueff et al.(2002)]{roueff2002} Roueff, E., Felenbok,
P., Black, J.~H., \& Gry, C.\ 2002, \aap, 384, 629

\bibitem[Rygl et al.(2011)]{rygl2011} Rygl, K.~L.~J., 
Brunthaler, A., Sanna, A., et al.\ 2011, arXiv:1111.7023 

\bibitem[Sakai et al.(2008)]{sea07} Sakai, N., Sakai, T., Hirota, T.
\& Yamamoto, S. 2008b, \apj, 672, 371

\bibitem[Sakai et al.(2009a)]{sea09} Sakai, N., Sakai, T., Hirota, T.,
Burton, M. \& Yamamoto, S. 2009a, \apj, 697, 769

\bibitem[Sakai et al.(2009b)]{sea09b} Sakai, N., Sakai, T., Hirota,
T., \& Yamamoto, S. 2009b, \apj, 702, 1025

\bibitem[Schilke et al.(1999)]{schilke1999} Schilke, P., Phillips,
T.~G., \& Mehringer, D.~M.\ 1999, The Physics and Chemistry of the
Interstellar Medium, 330 

\bibitem[Schmuttenmaer et al.(1990)]{schmuttenmaer1990} Schmuttenmaer, 
C.~A., Cohen, R.~C., Pugliano, N., Heath, J.~R., 
\& Cooksy, A.~L.\ 1990, Science, 249, 897 

\bibitem[Schneider et al.(2010)]{schneider2010} Schneider, N., Csengeri, T., Bontemps, S., et al.\ 2010, \aap, 520, A49 




\bibitem[Van Orden et al.(1995)]{vanorden1995} Van Orden, A., Cruzan,
J. D., Provencal, R. A., Giesen, T. F., et al.  1995, in ASP Conf.
Ser. 73, Proc. Airborne Astronomy Symp. on the Galactic Ecosystem, ed.
M. R. Haas, J. A.  Davidson, \& E. F. Erickson (San Francisco: ASP),
67

\bibitem[White et al.(2010)]{white2010} White, G.~J., Abergel, A.,
Spencer, L., et al.\ 2010, \aap, 518, L114 

\bibitem[Wilson \& Mauersberger(1990)]{wilson1990} Wilson, T.~L., \&
Mauersberger, R.\ 1990, \aap, 239, 305 


\bibitem[Zapata et al.(2012)]{zapata2012} Zapata, L.~A., Loinard, L.,
Su, Y.-N., et al.\ 2012, \apj, 744, 86 


\end{thebibliography}
\end{document}